\documentclass[]{spie}  %>>> use for US letter paper
\usepackage{amsmath,amsfonts,amssymb}
\usepackage{graphicx}
\usepackage[colorlinks=true, allcolors=blue]{hyperref}
\usepackage{braket}

\title{Optimal mode-sorting coronagraphy: limits of single-moded measurements for the Habitable Worlds Observatory}

\author[a]{Yinzi Xin}
\author[a]{Sebastiaan Y. Haffert}
\author[a]{Rico Landman}

\affil[a]{Sterrewacht Leiden, PO Box 9513, Niels Bohrweg 2, Leiden, The Netherlands}

\authorinfo{Further author information: (Send correspondence to Y.X.)\\Y.X.: E-mail: xin@strw.leidenuniv.nl}

\begin{document} 
\maketitle

\begin{abstract}
Conventional coronagraph architectures struggle to reach the theoretical limit of exoplanet detection at close working angles. Spatial mode-sorting is capable of reaching the theoretical limit, which previous work has calculated for a completely unresolved star whose signal lies entirely in the fundamental piston mode of the telescope. More recently, we have calculated optimized nulling modes as a function of the size of the star and planet parameters, with the goal of improving coronagraphic performance at $\sim \lambda/D$ working angles given partially resolved stars and complex telescope apertures. In this work, we further explore the limits of a coronagraph involving the measurement of a single spatial mode, with potential applications for the infrared arm of the Habitable Worlds Observatory. We present the achievable single-channel $1 \,\ \lambda/D$ planet throughput as a function of the single-channel stellar suppression (for a $0.03 \,\ \lambda/D$ diameter star), and show that the ability to measure a single optimized spatial mode can allow us to reach at least $\sim80\%$ of the theoretical information limit.
\end{abstract}

\keywords{optimal coronagraphy, exoplanets, nulling interferometry, quantum information theory}

\section{INTRODUCTION}
\label{sec:intro}  % \label{} allows reference to this section

Direct imaging of exoplanets is limited by the large brightness contrast between planets and their host stars and by the small angular separations at which planets appear. These challenges are particularly important for future observatories such as the ELT and Habitable Worlds Observatory (HWO), which aim to characterize potentially habitable planets through spectroscopy. Achieving the required levels of stellar suppression while maintaining planet throughput near $1\,\lambda/D$ remains a major technological challenge, as conventional coronagraph architectures typically lose sensitivity at these close-in separations \cite{kenworthy_high-contrast_2025}.

Recent work has explored exoplanet detection from the perspective of quantum-limited sensing, using tools from quantum information theory to identify coronagraph designs that approach the fundamental limits imposed by physics \cite{deshler_quantum_2026}. In this framework, coronagraphs can be viewed as devices that project the incoming electric field onto a chosen set of spatial modes. While conventional imagers perform this measurement using detector pixels, fiber-based nullers \cite{ruane_efficient_2018,echeverri_vortex_2019, por_single-mode_2020, haffert_single-mode_2020} and photonic lanterns \cite{xin_efficient_2022} instead project the field onto guided optical modes. Additionally, technologies based on interferometric meshes \cite{sirbu_astropic_2024} or multi-plane light converters \cite{morizur_programmable_2010, labroille_efficient_2014} can in theory implement arbitrary unitary transformations to demultiplex the incoming light on a user-defined set of spatial modes.

The measurement paradigm of spatial mode demultiplexing (SPADE) \cite{tsang_quantum_2016} has attracted interest for its superresolution capabilities. Although conventional coronagraphs, fiber- or lantern-based coronagraphs, interferometric meshes, and multi-plane light converters are all describable within the SPADE paradigm, the quantum-information perspective provided by SPADE gives us a powerful framework for analyzing and optimizing high-contrast imaging systems \cite{huang_ultimate_2023, deshler_quantum_2026}.

Initial work in Ref. \citenum{deshler_quantum_2026} showed that for an unresolved star, and in the limit of high planet contrast, removing the fundamental mode of the telescope results in a quantum-optimal coronagraph. Subsequent work in Ref. \citenum{xin_quantum-optimal_2026} extended the quantum measurement formalism to analyze resolved stars, and argued that calculating the measurement that maximizes information about the planet brightness provides the most generally useful coronagraph, especially for science cases relying on measurements of exoplanet spectra. Ref. \citenum{xin_quantum-optimal_2026} calculated this optimal measurement for several example science cases, including a hypothetical coronagraph for HWO for spectrally characterizing a habitable-zone exoplanet at infrared wavelengths, provided the position of the planet can be identified beforehand using visible wavelength coronagraphy. In this proceedings, we summarize the theoretical approach of Ref. \citenum{xin_quantum-optimal_2026} to calculating this optimal measurement, and we also present additional calculations pertaining to the HWO infrared coronagraph example.

\section{SUMMARY OF THEORETICAL APPROACH}

Ref. \citenum{xin_quantum-optimal_2026} contains a primer on quantum mechanics notation and an explanation of how this notation is related to mathematical descriptions of the electric field conventionally used in high contrast imaging instrumentation. Following that work, we write the density matrix for finite-size stars with a single off-axis planet as

\begin{equation} \label{eq:density_matrix}
    \hat{\rho} = \frac{(1-c)}{N_s}\sum_i\ket{\psi_{s_i}}\bra{\psi_{s_i}}+c \ket{\psi_{p}}\bra{\psi_{p}}.
\end{equation}

Here, $\ket{\psi_{s_i}}$ describe the states of the incoherent sources that make up the disk of the star, $N_s$ is a parameter that normalizes the flux of the star based on the number of states used to sample the disk, $\ket{\psi_{p}}$ describes the state corresponding to hte planet, and $c$ describes the relative flux contribution of the planet to the whole system (which is approximately its flux ratio in the limit of $c\rightarrow0$). The symmetric logarithmic derivative (SLD) for estimating $c$ is the most sensitive measurement possible, achieving the supremum of classical Fisher information (CFI), known as the quantum Fisher information (QFI). The SLD is denoted as $\hat{L}_c$ and given by

\begin{equation}
    \partial_c \hat{\rho}(c) = \frac{1}{2}(\hat{L}_c \hat{\rho}(c) + \hat{\rho}(c)\hat{L}_c).
\end{equation}

In the basis that diagonalizes the density matrix given by $\hat{\rho} = \sum_m \zeta_m \ket{\zeta_m} \bra{\zeta_m}$, the SLD can be calculated as

\begin{equation} \label{eq:sld_closed_form}
    \hat{L}_c = 2 \sum_{m,n} \frac{\braket{\zeta_n|\partial_c\hat{\rho}|\zeta_m}}{\zeta_m+\zeta_n} \ket{\zeta_n}\bra{\zeta_m}.
\end{equation}

\section{PROSPECTS FOR A SINGLE-MODED INFRARED CORONAGRAPH FOR HWO}

Ref. \citenum{xin_quantum-optimal_2026} calculated the SLD for single planet of known location and showed that, depending on the parameters of the system, typically $>80\%$ of the QFI is concentrated in a single spatial mode. In this section, we further explore the prospect of an infrared coronagraph for HWO consisting of the measurement of a single spatial mode, assuming that the planet location (especially position angle) can be constrained beforehand with visible wavelength imaging.

\begin{figure*}[t]
\begin{center}
	\includegraphics[scale = 0.5]{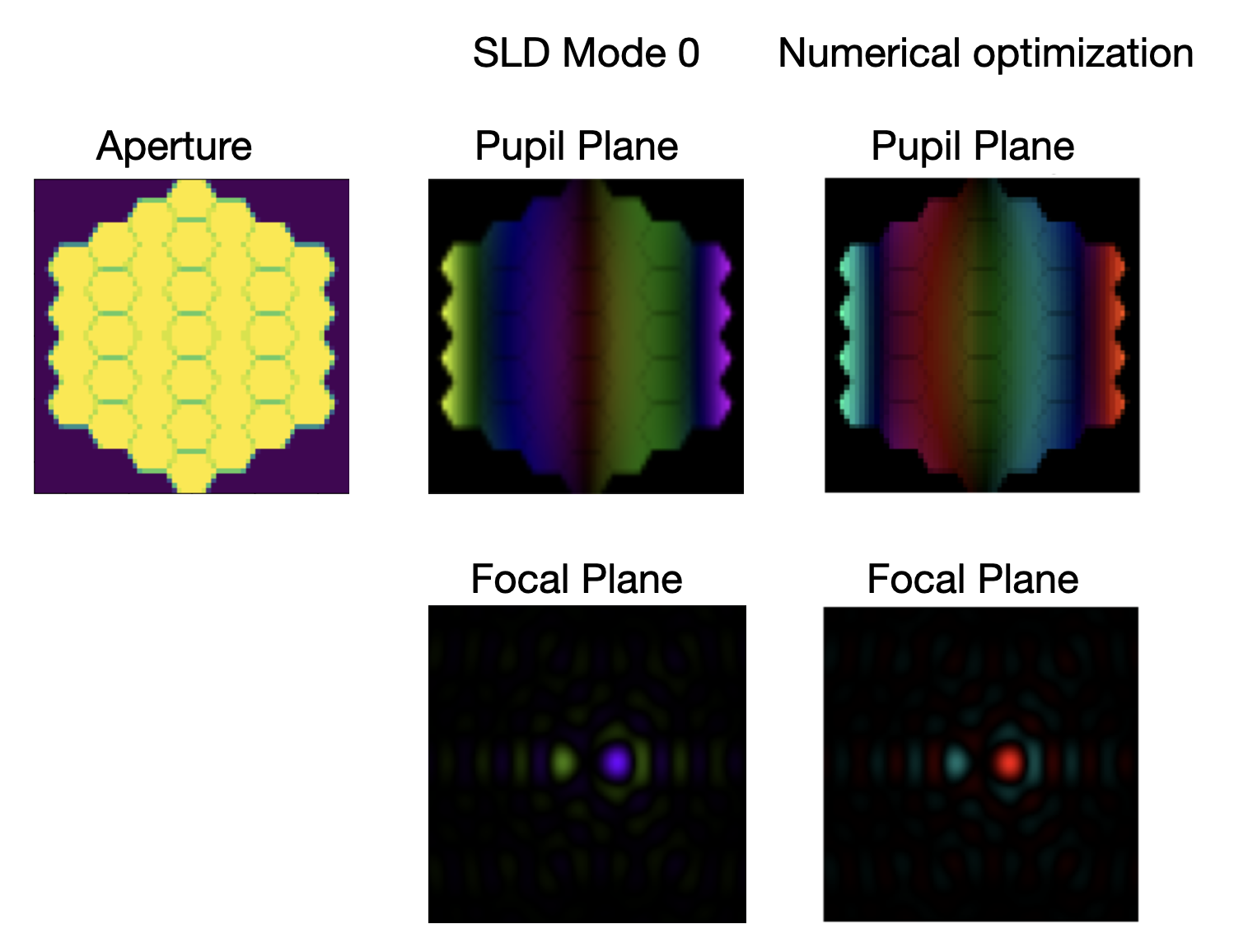}
	\caption{\label{fig:example_results} Left) The aperture assumed in these simulations. Furthermore, the stellar radius is assumed to be $0.03 \,\ \lambda/D$, and the planet is assumed to be at $1 \,\ \lambda/D$ in the positive X direction. Middle) The single-moded measurement shown in the pupil plane (top) and focal plane (bottom) obtained by using the highest information mode from the calculation of the SLD, denoted here as SLD Mode 0. Right) The single-moded measurement shown in the pupil and focal planes as calculated using a classical numerical optimization. Pupil plane extents are from -0.5 to 0.5 (with an aperture diameter normalized to 1), while focal plane extents are from -5 $\lambda/D$ to $5 \lambda/D$. Both of these methods are described in Ref. \citenum{xin_quantum-optimal_2026}, and achieve $\eta_s \approx 1\times10^{-9}$, with $\eta_p$ at $1 \,\ \lambda/D$ of approximately $19\%$, and a maximum $\eta_p$ of $73\%$ at $1.8 \,\ \lambda/D$.}
\end{center}
\end{figure*}

\begin{figure*}[t]
\begin{center}
	\includegraphics[scale = 0.56]{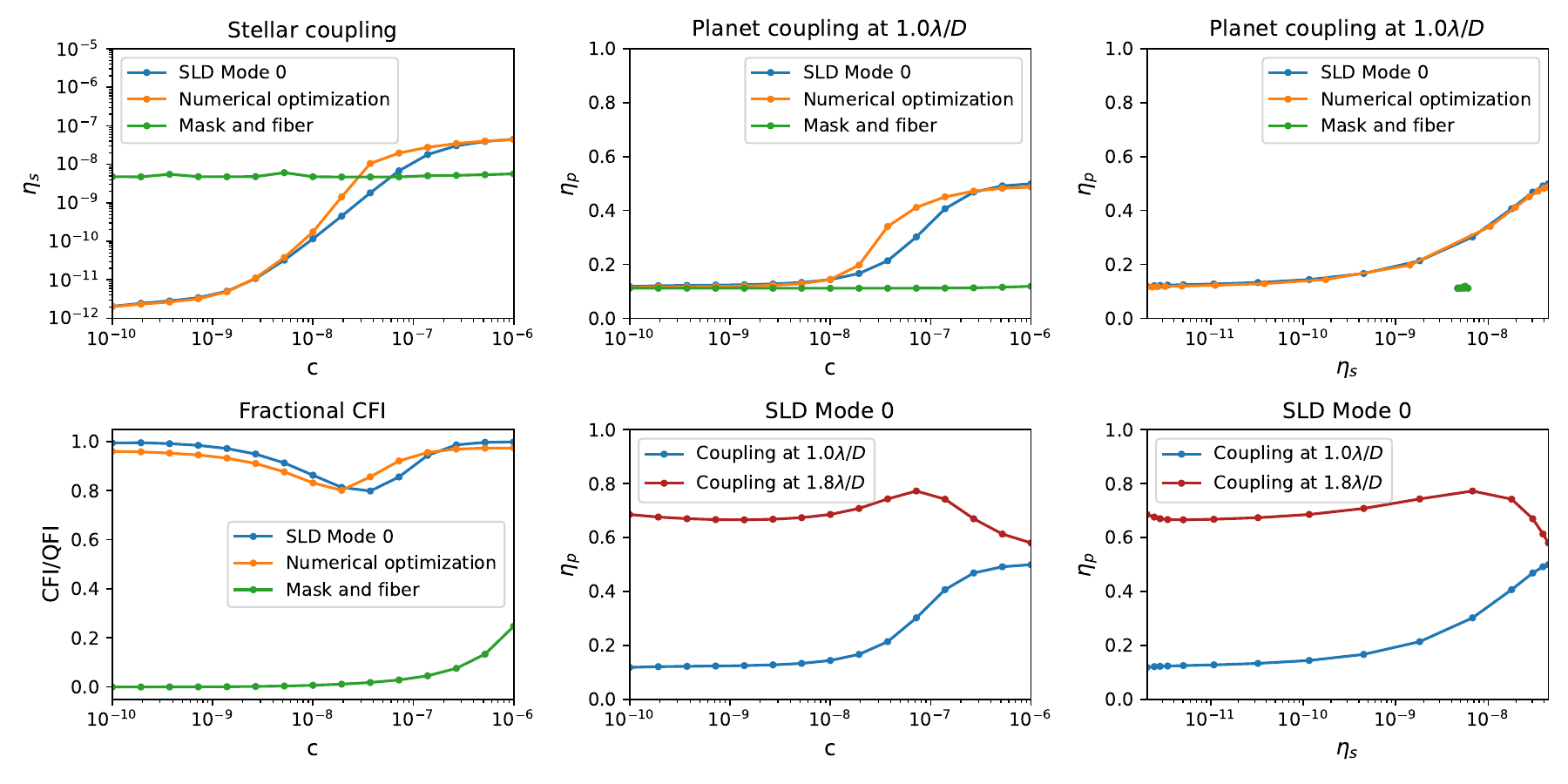}
	\caption{\label{fig:c_sweep} Top left) Stellar coupling in a single-moded measurement as a function of the parameter $c$. The stellar radius was assumed to be $0.03 \lambda/D$. Top middle) Planet coupling (at the design separation of $1 \,\ \lambda/D$) in a single-moded measurement for the same system as a function of the parameter $c$. Top right) The planet coupling at $1 \,\ \lambda/D$ as a function of the stellar coupling in the mode. The spread in results for the mask and fiber architecture is due to scatter in the numerical optimization process. Bottom left) The CFI achievable with a single-moded measurement, divided by the QFI, as a function of the design value of $c$. The ability to sort an arbitrary spatial mode provides an enormous performance boost relative to, for example, a numerically optimized fiber nuller. Bottom middle and right) The planet coupling achieved by the highest information mode of the SLD at 1.8 $\lambda/D$ compared to that at 1.0 $\lambda/D$, as a function of the design value of $c$ and the stellar throughput $\eta_s$ respectively. While these modes are optimized for a planet location of 1.0 $\lambda/D$, the physical constraints due to the overlap of the resolved star with close-in planet signals result in solutions that typically have higher planet throughput further out.}
\end{center}
\end{figure*}

We calculate the spatial profile of this single-moded measurement using two different methods, both discussed in more detail in Ref. \citenum{xin_quantum-optimal_2026}: 1) directly using the highest information mode from the SLD and 2) a semi-analytic optimization that maximizes the S/N within a single mode. An example result for the same system calculated using each method is shown in Figure \ref{fig:example_results}. The stellar throughput $\eta_s$ and planet throughput $\eta_p$ (at $1\,\ \lambda/D$) as a function of the design flux ratio $c$ are shown in the top left and top middle panels of Figure \ref{fig:c_sweep}.

While the two results for a given value of the planet flux ratio $c$ may differ slightly, we show in the top right of Figure \ref{fig:c_sweep} that the achieved planet throughput ($\eta_p$) as a function of the stellar rejection ($\eta_s$) is the roughly the same for both methods. Note that we have designed the coronagraph to target a separation of $1 \,\ \lambda/D$, where the physical overlap between the planet signal and the resolved stellar signal heavily impacts the achievable planet throughput given a high level of stellar suppression. However, as shown in the bottom middle and bottom right panels of Figure \ref{fig:c_sweep}, the planet throughput over all space for these measurements peaks further out, and can reach $\sim70\%$ at $\sim 1.8 \lambda/D$ given $\eta_s \sim 10^{-9}$.

Because of the presence of exozodiacal dust not accounted for in the measurement problem, only stellar suppression levels of $\sim 10^{-9}$ are expected to be practical for HWO's infrared coronagraph \cite{ertel_hosts_2020, ertel_review_2025, stark_paths_2024}, which can be achieved using a design $c$ parameter of $\sim 10^{-8}$. We show in the bottom left of Figure \ref{fig:c_sweep} that, somewhat unfortunately, the fractional QFI of the optimal single-mode measurement with the design $c\sim10^{-8}$ is near the minimum of $80\%$. However, the sorting of a single spatial mode (as opposed to one involving multiple modes) would be significantly easier to manufacture, and it would require only one spectrograph trace, reducing the overall complexity of the instrument.

Additionally, an optimal single-moded coronagraph that achieves $80\%$ of the QFI would still outperform most other potential coronagraph designs. For example, we can examine the fiber nuller coronagraph architecture wherein a complex mask in the pupil plane modifies the beam before it is injected into the fiber (see examples in Ref. \citenum{haguenauer_deep_2006, ruane_efficient_2018}, as well as Chapter 21 of Ref. \citenum{jovanovic_2023_2023} for a review). However, rather than using an a priori phase pattern such as a piston shift or a vortex, we have used an autodifferentiable optical model built in dLux \cite{desdoigts_differentiable_2023} to jointly optimize the complex mask, the fiber location, and the fiber mode-field diameter to maximize the CFI of the planet signal for various values of $c$. This CFI ($F_c(c)$) cost function can be written as:

\begin{equation}
    F_c(c) = \frac{(\eta_p-\eta_s)^2}{(1-c)\eta_s+c\eta_p}.
\end{equation}

\begin{figure*}[t]
\begin{center}
	\includegraphics[scale = 1.0]{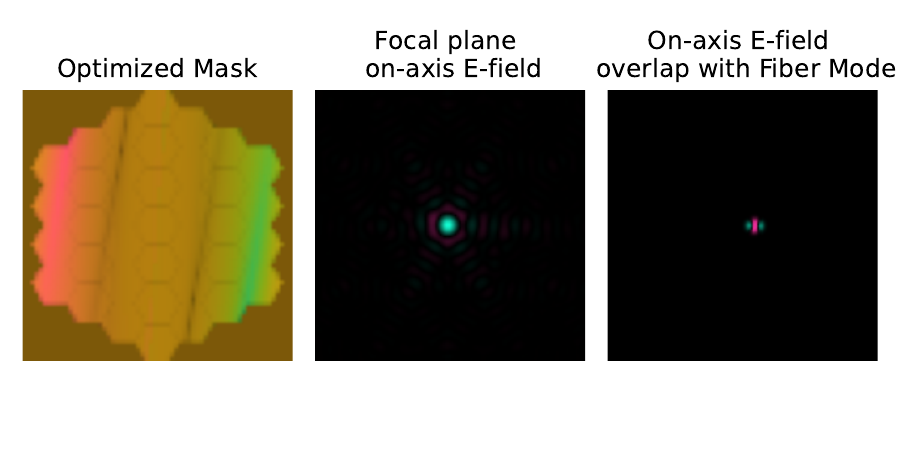}
	\caption{\label{fig:fiber_nulling_example} Left) The complex pupil plane mask of a fiber nuller optimized to maximize the Fisher information of a faint off-axis planet at $1 \,\ \lambda/D$, given a stellar radius of $0.03 \,\ \lambda/D$. The extent is from -0.5 to 0.5, with aperture diameter normalized to 1. The dark diagonal streaks are numerical artifacts that do not noticeably impact the performance of the fiber nuller. Middle) The on-axis electric field in the focal plane, after transmission through the complex pupil mask. The extent is from -15 $\lambda/D$ to 15 $\lambda/D$. Right) The overlap of the on-axis electric field in the focal plane with the optimal gaussian fiber mode, showing that the complex mask has modified the beam of the starlight to create a null at the location of the fiber. The performance metrics characterizing this fiber nuller are shown in Figure \ref{fig:c_sweep}. The results do not vary significantly with the value of $c$ used in the cost function.}
\end{center}
\end{figure*}

We present the results of this optimization in Figure \ref{fig:fiber_nulling_example}. Figure \ref{fig:c_sweep} shows that the performance of even this optimal fiber nuller is far from that achievable by measuring a single optimal spatial mode. The addition of phase-induced amplitude apodizing optics \cite{guyon_phase-induced_2003} may slightly improve performance, but only slightly, and still would leave a large gap to be bridged.

\section{OUTLOOK AND CONCLUSION}

Other proposed instrument concepts for HWO that generate one nulled output using photonic lanterns \cite{morsy_combining_2026} and/or photonic circuits \cite{sirbu_astropic_2024} would also be implementations of a single-moded measurement; however, it remains to be seen how closely these architectures can approach the single-mode information limit given a finite number of samples and beam-combinations. A multi-plane light converter \cite{morizur_programmable_2010, labroille_efficient_2014} could be designed to directly sort the optimal spatial mode, but the manufacturability of such a device, at high enough fidelity to reach HWO contrast levels, is also currently an open question.

We recommend advancing all potential mode-sorting technologies, as each architecture will face a different, not yet fully known, set of practical limitations. Ultimately, while these technologies are all capable of arbitrary mode-sorting, significant effort is still needed for any of them to achieve deep, broadband nulls with high throughput, high stability, and desirable polarization properties. However, advancing these technologies would open up a new regime of coronagraphic performance at small working angles, with the potential to dramatically increase the science yield of HWO.

\acknowledgments % equivalent to \section*{ACKNOWLEDGMENTS}
The authors acknowledge support from NWO Award 184.036.004.

% References
\bibliography{report} % bibliography data in report.bib
\bibliographystyle{spiebib} % makes bibtex use spiebib.bst

\end{document}